\documentclass[24pt]{article}
\usepackage{amssymb}
\usepackage{amsfonts}
\usepackage{amssymb,amsmath}
\usepackage{mathrsfs}
\usepackage{latexsym}
\usepackage{amsmath}
\usepackage{latexsym}
\usepackage[cp1251]{inputenc}
\usepackage{graphicx}
\usepackage{color}

\title{Application of a quantum wave impedance method for study of infinite and semi-infinite periodic media}
\author{O. I. Hryhorchak\\
{\small Department for Theoretical Physics, Ivan Franko National
University of Lviv,}\\
{\small 12, Drahomanov Str., Lviv, UA--79005,
Ukraine}\\
\small{\it{Orest.Hryhorchak@lnu.edu.ua}}}

\def\ch{\mathop{\rm ch}\nolimits}
\def\sh{\mathop{\rm sh}\nolimits}
\def\th{\mathop{\rm th}\nolimits}

\def\tg{\mathop{\rm tg}\nolimits}
\def\ctg{\mathop{\rm ctg}\nolimits}

\begin{document}
\renewcommand{\abstractname}{Abstract}
\maketitle

\begin{abstract}
This paper is dedicated to an application of a quantum wave impedance approach for a study of infinite and semi-infinite periodic systems. Both a Dirac comb and a $\delta-\delta'$ comb as well as a Kronig-Penney model are considered. It was shown how to reformulate the problem of an investigation of mentioned systems in terms of a quantum wave impedance and it was demonstrated how much a quantum wave impedance approach simplifies studying these systems compared to other methods. The illustation of such a simplification was provided by application of classical approach, transfer matrix technique  and a quatum wave impedance method  for solving Kronig-Penney model. 
\end{abstract}

\section{Introduction}
In 1990 Hague and co-authors \cite{Hague_Hague_Khan:1990} with the help of a quantum wave impedance concept and an analogy with an electrical transmission line, deve\-loped a technique which allows calculating the energy bands of periodic potential structures. They also showed that periodic structures with a non-rectangular potential distribution can be analysed using a quantum wave impedance approach. Authors considered periodic sys\-tems with different potential distributions and in fact sketched the method for the calculation of eigenstates of a periodic potential structure with an arbitrary potential variation within the unit cell. Later Nelin and co-authors applied the impedance method for crystal-like structures, namely, electromagnetic, phonon and photon crystals \cite{Nelin:2004, Nelin:2005, Nelin:2006, Nelin:2009_2, Khatyan_Gindikina_Nelin:2015, Nazarko_etall:2009, Gindikina_Zinger_Nelin:2015, Nelin:2007_1, Nelin_Nazarko:2012, Babushkin_Nelin:2011, Nazarko_etall:2011, Nelin:2012, Nazarko_etall:2011_1, Nazarko_Timofeeva_Nelin:2010, Nelin_Zinher_Popsui:2017, Nelin:2004_1, Nazarko_etall:2015} and described a zone diagram formation in these periodic systems. It was shown that the results of these papers can be directly used for a design of nano-electronic devices, in particular, filters. In \cite{Nazarko_etall:2015} the filter based on
electromagnetic crystal inhomogeneities is studied and the structure of a narrowband filter based on the scheme of a Fabry-Perot resonator and amplitude-frequency characteristics of the filter are described. In \cite{Nazarko_Timofeeva_Nelin:2010} the interaction of an electromagnetic field with the different inhomogeneities of electromagnetic crystals is described. In
\cite{Khatyan_Gindikina_Nelin:2015} the superlattice structures were investigated by an impedance method and the input impedance characteristics of these structures were studied as well as the patterns of their band diagram formation.

Using a 1D periodic lattice as a model of crystals (see, for example \cite{Ashcroft_Mermin:1976}) we assume the equality of a distance between positive ions which cause the potential inside a crystal. Thus, we have a quantum mechanical problem of electron which is subjected to a regular potential. Even in a case of omitting an electron-electron interaction this problem is exactly solvable only for some forms of a periodic potential. But exactly solvable models are very important, in particular because they are a good illustration of how a zone structure of solids forms. In this paper we are going to show how an approach of a quantum wave impedance can be applied to the description of 1D periodic media and what advantages are of using this method compared to well-known ones.

\section{Quantum wave impedance for an infinite periodic systems}

Bloch-Floquet theorem \cite{Floquet:1883, Bloch:1929} states that for an infinite periodic potential $U(x)$ with a period $L$, that is $U(x+L)=U(x)$, a wave function $\psi(x)$ of such a system has a property as follows:
\begin{eqnarray}\label{psi_exp_fi}
\psi(x+L)=\exp(ikx)\phi(x),
\end{eqnarray}
where $k$ is a quasi wave-vector and $\phi(x)$ is a periodic function with a period $L$: $\phi(x+L)=\phi(x)$.
On the base of formula which relates a quantum wave impedance function and a wave function \cite{Arx1:2020, Arx2:2020} we get an expression for a quantum wave impedance in a case of a periodic potential:
\begin{eqnarray}
Z(x)=\frac{\hbar k L}{m}+\frac{\hbar}{im}\frac{\phi'(x)}{\phi(x)}.
\end{eqnarray}
Integrating both sides of the previous equation within elementary cell we get
\begin{eqnarray}\label{Z_periodic}
\int_0^L Z(x) dx = \frac{\hbar \gamma L}{m}+\frac{\hbar}{im}\ln\frac{\phi(L)}{\phi(0)}.
\end{eqnarray}
Because of a periodicity of a function $\phi(x)$ we have $\phi(0)=\phi(L)$ and
\begin{eqnarray}
\frac{im}{\hbar}\int_0^L Z(x) dx = i\gamma L. 
\end{eqnarray}
Additionally to this formula we should use a periodic and/or a matching condition for a quantum wave impedance. In a case of non-singular potentials these conditions are as follows
\begin{eqnarray}\label{Z_cond_pm}
Z(x_i)=Z(x_i+L), \quad Z(x_i+0)=Z(x_i-0),\quad \forall x_i \in (0\ldots L).
\end{eqnarray} 
In a case of zero-range singular potentials these conditions depend on the type of a singularity. We will illustrate it in the next sections.

This simple formula (\ref{Z_periodic}) with conditions (\ref{Z_cond_pm}) allows finding a dispersion relation for numbers of infinite periodic systems. We will consider the most known of them below.

\section {Infinite Dirac comb}

A periodic potential which is formed by the periodically placed $\delta$-functions is one of the simplest and most widely used models. With its help one models ideal lattices, influence of defects and edges in crystals, surface states etc.
A periodic potential of a Dirac comb has the following form
\begin{eqnarray}
U(x)=\sum_{n=-\infty}^\infty  \alpha\delta(x-nL).
\end{eqnarray}
The solution of the equation for a quantum wave impedance with this potential is as follows
\begin{eqnarray}
Z(x)=z_0\th(ik_0x+\phi),\qquad k_0=\frac{\sqrt{2mE}}{\hbar}.
\end{eqnarray}
Using formula (\ref{Z_periodic}) we get 
\begin{eqnarray}
\frac{\ch(ik_0L+\phi)}{\ch(\phi)}=\exp[i\gamma L].
\end{eqnarray}
Taking a real part of both sides of the previous equation we get   
\begin{eqnarray}
\cos(k L)=\cos(k_0L)-\sin(k_0L)\Im\left[\th(\phi)\right],
\end{eqnarray}
where $k$ is a Bloch's quasi wave-vector. Now our task is to calculate the value for $\Im[\th(\phi)]$. On the base of both a periodic ($Z(x+L)=Z(x)$) and a matching condition for a $\delta$-potential \cite{Arx4:2020} we find that  
\begin{eqnarray}
Z(a-0)-Z(+0)=i\frac{2\alpha}{\hbar},
\end{eqnarray}
which is
\begin{eqnarray}
\th(ik_0L+\phi)-\th(\phi)=i\frac{2\alpha}{z_0\hbar},
\end{eqnarray}
where $z_0=\hbar k_0/m$. We can rewrite it in the following way
\begin{eqnarray}
\th[ik_0L]-\th[ik_0L]\th^2[\phi]=\frac{2i\alpha}{z_0\hbar}(1+\th[ik_0L]\th[\phi]).
\end{eqnarray} 
Taking a $\Re$ part from both sides of this relation we get:
\begin{eqnarray}
2\tg(k_0L)\Im(\th[\phi])\Re(\th[\phi])
=-\frac{2\alpha}{z_0\hbar}\tg(k_0L)\Re(\th[\phi]).
\end{eqnarray}
Similarly, taking an $\Im$ part we obtain:
\begin{eqnarray}
\tg(k_0L)(1+\Im^2(\th[\phi])-\Re^2(\th[\phi]))=
\frac{2\alpha}{z_0\hbar}\left(1-\tg[k_0L]\Im(\th[\phi])\right).
\end{eqnarray} 
From this system of equations we have that 
\begin{eqnarray}
\Im(\th(\phi))&=&-\frac{\alpha}{z_0\hbar},\nonumber\\
\Re(\th(\phi))&=&\pm\sqrt{1-\frac{\alpha^2}{z_0^2\hbar^2}
	-\frac{2\alpha}{z_0\hbar}\ctg(k_0L)}
\end{eqnarray}
and finally
\begin{eqnarray}
\cos(k L)=\cos(k_0L)+\frac{m\alpha }{k_0\hbar^2}\sin(k_0 L)
\end{eqnarray}
or after introducing the following notations
\begin{eqnarray}
\xi=k_0L,\qquad p=\frac{m\alpha L}{\hbar^2}
\end{eqnarray}
we get the same relation in a more familiar form
\begin{eqnarray}\label{dr_DC}
\cos(k L)=\cos(\xi)+\frac{p}{\xi}\sin(\xi).
\end{eqnarray}
It is a well-known dispersion relation for a Dirac comb \cite{Cordoba:1989}.

%Дозволені зони --- це це зони власних значень енергії %кристала.

\section{Wave functions of a Dirac comb }
After the dispersion relation is found the next question is about wave functions of a Dirac comb. Consider the region $0<x<L$. The expression for a quantum wave impedance in this region we have obtained in the previous section
\begin{eqnarray}
Z(x)=z_0\th[ik_0L+\phi].
\end{eqnarray}
Reminding the relation between a quantum wave impedance function and a wave function \cite{Arx1:2020} we get that
\begin{eqnarray}
\!\!\!\!\psi(x)=\exp\left(\frac{im}{\hbar}\int  Z(x)dx\right)=A\left(\frac{}{}\exp[ik_0x]+r\exp[-ik_0x]\right)\!\!, 
\end{eqnarray}
where $r=\exp[-2\phi]$, and
\begin{eqnarray}
\frac{\ch[ik_0L+\phi]}{\ch[\phi]}&=&
\frac{\ch(ik_0L)(1+r)+\sh(ik_0L)(1-r)}{1+r}=
\nonumber\\
&=&\frac{\exp[ik_0L]+r\exp[-ik_0L]}{1+r}=\exp[ik L],
\end{eqnarray}
\begin{eqnarray}
r=\frac{\exp[ik_0L]-\exp[ik L]}{\exp[ik L]-\exp[-ik_0L]}=
\frac{\exp[i(k_0-k)L]-1}{1-\exp[-i(k_0+k)L]}.
\end{eqnarray}
Thus, we obtained an expression for a wave function $\psi_0(x)$ in the region: $0<x<a$. And what about the rest space. To answer this question let's remind an expression (\ref{psi_exp_fi}) which in our case gives
\begin{eqnarray}
\phi(x)=\psi_0(x)\exp[-ik x].
\end{eqnarray}
After an introducing of $x=x'+nL$ we can write
\begin{eqnarray}
\psi(x'+nL)=\phi(x'+nL)\exp[ik(x'+nL)]= \phi(x')\exp[ik x']\exp[ik nL].
\end{eqnarray}
Taking into account that
\begin{eqnarray}
\phi(x')=\psi_0(x')\exp[ik x']
\end{eqnarray}
we get a wave function for a whole Dirac comb
\begin{eqnarray}
\psi(x)=\psi_0(x-nL)\exp[ik nL]\quad (na\leq x\leq(n+1)L).
\end{eqnarray}
The other way to the same result runs through a direct using of a quantum wave impedance. So let's write a series of simple transformations
\begin{eqnarray}
&&\frac{\ch[ik_0(n+1)L+\phi]}{\ch[ik_0anL+\phi]}=\frac{\ch(ik_0(n+1)L)(1+r)+\sh(ik_0(n+1)L)(1-r)}
{\ch(ik_0nL)(1+r)+\sh(ik_0nL)(1-r)}=\nonumber\\
&&=\frac{\exp[ik_0(n+1)L]+r\exp[-ik_0(n+1)L]}{\exp[ik_0nL]+r\exp[-ik_0nL]}=\frac{\exp[ik_0L]+r\exp[-2ik_0nL]\exp[-ik_0L]}{1+r\exp[-2ik_0nL]}.\nonumber\\
\end{eqnarray}
Then
\begin{eqnarray}
\frac{\exp[ik_0L]+r\exp[-2ik_0nL]\exp[-ik_0L]}{1+r\exp[-2ik_0nL]}=\exp[ik L].
\end{eqnarray}
Consequently
\begin{eqnarray}
r\exp[-2ik_0nL]=\frac{\exp[ik_0L]-\exp[ik L]}{\exp[ik L]-\exp[-ik_0L]}=\frac{\exp[i(k_0-k)L]-1}{1-\exp[-i(k_0+k)L]}=r_0
\end{eqnarray}
and
\begin{eqnarray}
\psi(x)&=&A\left(\exp[ik_0x]+r_0\exp[2ik_0nL]\exp[-ik_0x]\right)=\nonumber\\
&=&A\exp[ik_0nL]\left(\exp[ik_0(x-nL)]+\right.\left.r\exp[-ik_0(x-nL)]\right),
\end{eqnarray}
which means that
\begin{eqnarray}
\psi(x)=\psi_0(x-nL)\exp[ik_0nL].
\end{eqnarray}
Finally, for a quantum wave impedance function $Z(x)$ in the whole region of a Dirac comb we have
\begin{eqnarray}\label{Z_DC_periodic}
Z(x)=z_0\frac{\exp[ik_0x]-r_0\exp[2ik_0nL]\exp[-ik_0x]}{\exp[ik_0x]-r_0\exp[2ik_0nL]\exp[-ik_0x]}=z_0\th[ik_0x+\phi(n)],
\end{eqnarray}
where
\begin{eqnarray}
\phi(n)=\phi_0-ik_0nL,\qquad \phi_0=-\frac{1}{2}\ln(r_0).
\end{eqnarray}
%\vspace{0.5cm}

\section{Infinite $\delta-\delta'$ comb}
One might notice that for the last decades there were  many papers dedicated to a problem of a single
$\delta-\delta'$-potential but there were no papers which studied a lattice of $\delta-\delta'$-potentials besides  preprint \cite{Gadella:2019}. 
So, in this section we consider the extended analogue of a Dirac comb, namely a $\delta-\delta'$ comb. The potential energy of this model has a form
\begin{eqnarray}
U(x)=\sum_{n=-\infty}^\infty\left\{ -\alpha\delta(x-nL)+\beta\delta'(x-nL)\right\}.
\end{eqnarray}
The solution of the equation for a quantum wave impedance function with this potential within the  elementary cell is the same as in a case of a Dirac comb
\begin{eqnarray}
Z(x)=\th(ik_0x+\phi),\qquad k_0=\frac{\sqrt{2mE}}{\hbar}.
\end{eqnarray}
The matching condition for a $\delta-\delta'$-potential we have found in the paper \cite{Arx4:2020}. Thus,
we get
\begin{eqnarray}\label{Z_per_dd'}
\exp[ik L]=\exp\left[\frac{im}{\hbar}\int\limits_0^L\th(ik_0x+\phi)dx +f0 \right]=\frac{\ch(ik_0L+\phi)}{\cos(\phi)}\exp[f_0], 
\end{eqnarray}
where, taking into account the equality $Z(x+L)=Z(x)$,
\begin{eqnarray}
f0=\lim_{\varepsilon\rightarrow 0}\frac{im}{\hbar}\int\limits_{0}^\varepsilon Z(x)+\frac{im}{\hbar}\int\limits_{a-\varepsilon}^a Z(x)=\frac{im}{\hbar}\int\limits_{-\varepsilon}^\varepsilon Z(x)=\ln\left[\frac{\psi(0+)}{\psi(0-)}\right]=\ln\left[\frac{1+\tilde{\beta}}{1-\tilde{\beta}}\right]\!.
\end{eqnarray}
Taking a real part of both sides of (\ref{Z_per_dd'})  we get
\begin{eqnarray}\label{disp_res_dd'_nf}
\cos(k L)=\frac{1+\tilde{\beta}}{1-\tilde{\beta}}\left\{\frac{}{}\cos(k_0L)-\sin(k_0L)\Im\left[\th(\phi)\right]\right\},
\end{eqnarray}
where $k$ is a Bloch's quasi wave-vector.
For a calculation of $\Im[\th(\phi)]$ we use a matching \cite{Arx4:2020} and a periodic condition $Z(L-0)=Z(0-)$. Consequently, we get that 
\begin{eqnarray}
Z(0+)=\frac{i\hbar\tilde{\alpha}}{m(1+\tilde{\beta})^2}+
\frac{(1-\tilde{\beta})^2}
{(1+\tilde{\beta})^2}Z(L-0)
\end{eqnarray}
or in an explicit form 
\begin{eqnarray}
z_0\th(\phi)=\frac{i\hbar\tilde{\alpha}}{m(1+\tilde{\beta})^2}+
\frac{(1-\tilde{\beta})^2}
{(1+\tilde{\beta})^2}z_0\th(ik_0L+\phi).
\end{eqnarray}
After simple transformations as follows 
\begin{eqnarray}
(1\!+\!\tilde{\beta})^2\left(\th[\phi]\!+\!i\tg[k_0L]\th^2[\phi]\right)\!=\!\frac{i\hbar\tilde{\alpha}}{m}(1\!+\!i\tg[k_0L]\th[\phi])\!+\!(1+\tilde{\beta})^2(i\tan[k_0L]\!+\!\th[\phi])
\end{eqnarray}
and taking $\Re$ part from both sides of the previous relation
\begin{eqnarray}
(1+\tilde{\beta})^2\Re(\th[\phi])(1-2\tg[k_0L]
\Im(\th[\phi]))
=-\frac{\hbar\tilde{\alpha}}{m}\tg[k_0L]\Re(\th[\phi])
+(1-\tilde{\beta})^2\Re(\th[\phi])
\end{eqnarray}
we obtain
\begin{eqnarray}
\Im(\th[\phi])=\frac{\hbar\tilde{\alpha}}{2z_0m(1+\tilde{\beta})^2}+\frac{2\tilde{\beta}}{\tan[k_0L](1+\tilde{\beta})^2}.
\end{eqnarray}
Substituting it into the expression (\ref{disp_res_dd'_nf}) we get:
\begin{eqnarray}
\cos(k L)=\frac{1+\tilde{\beta}^2}{1-\tilde{\beta}^2}\left\{\frac{}{}\cos(k_0L)-\frac{\hbar\tilde{\alpha}}{2z_0m(1+\tilde{\beta}^2)}\sin(k_0L)\right\}.
\end{eqnarray}
Reminding that $\xi=kL$ and $p=mL\alpha/\hbar$
we finally obtain
\begin{eqnarray}
\cos(k L)=\frac{1+\tilde{\beta}^2}{1-\tilde{\beta}^2}
\left\{\frac{}{}\cos(\xi)-
\frac{p}{\xi(1+\tilde{\beta})^2}\sin(\xi)\right\}.
\end{eqnarray}
We got the dispersion relation for a $\delta-\delta'$ comb and it is the same as in the mentioned preprint \cite{Gadella:2019}. If we assume $\tilde{\beta}=0$ then our expression reduces to the one for a Dirac comb.

\begin{figure}[h!]
	\centerline{
		\includegraphics[clip,scale=0.86]{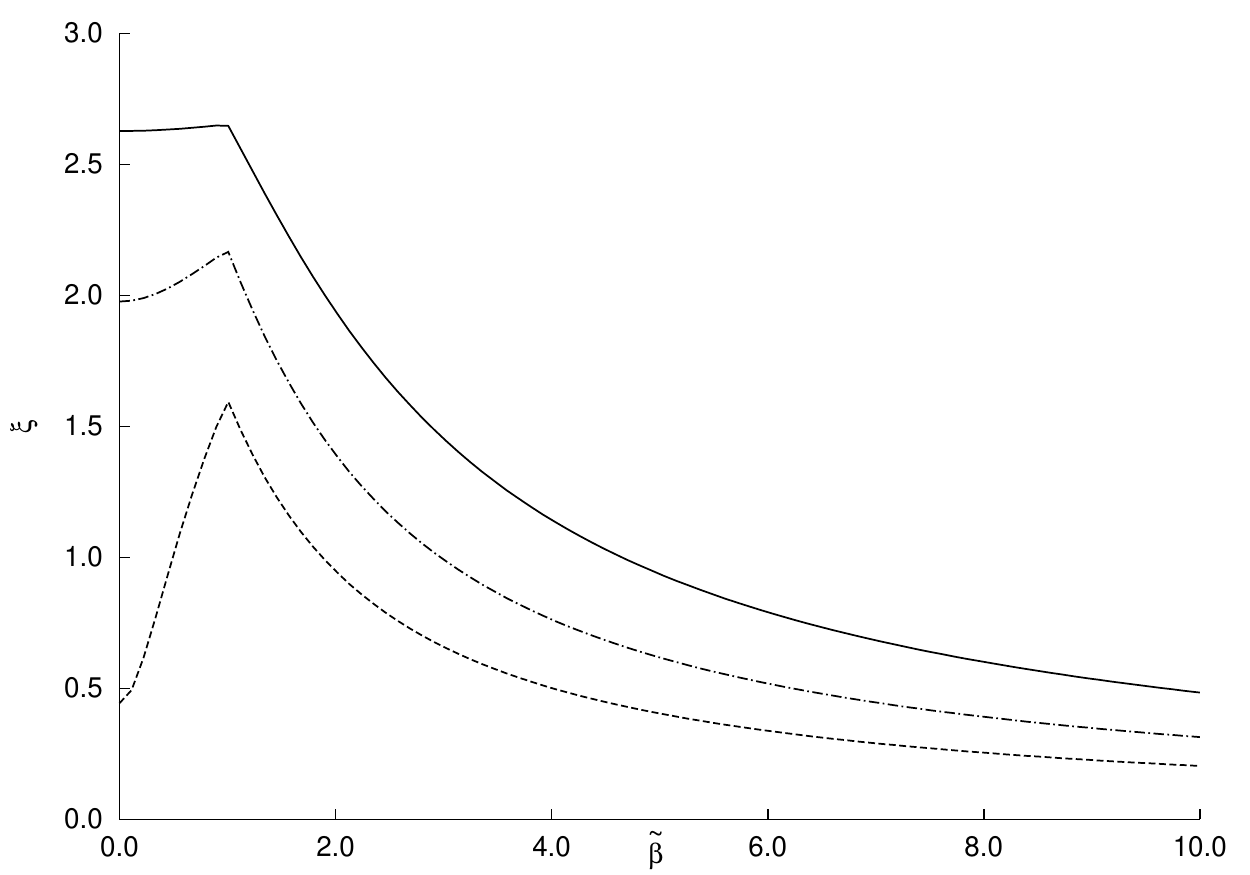}}
	\vspace{-0.3cm}
	\centerline{
		\includegraphics[clip,scale=0.86]{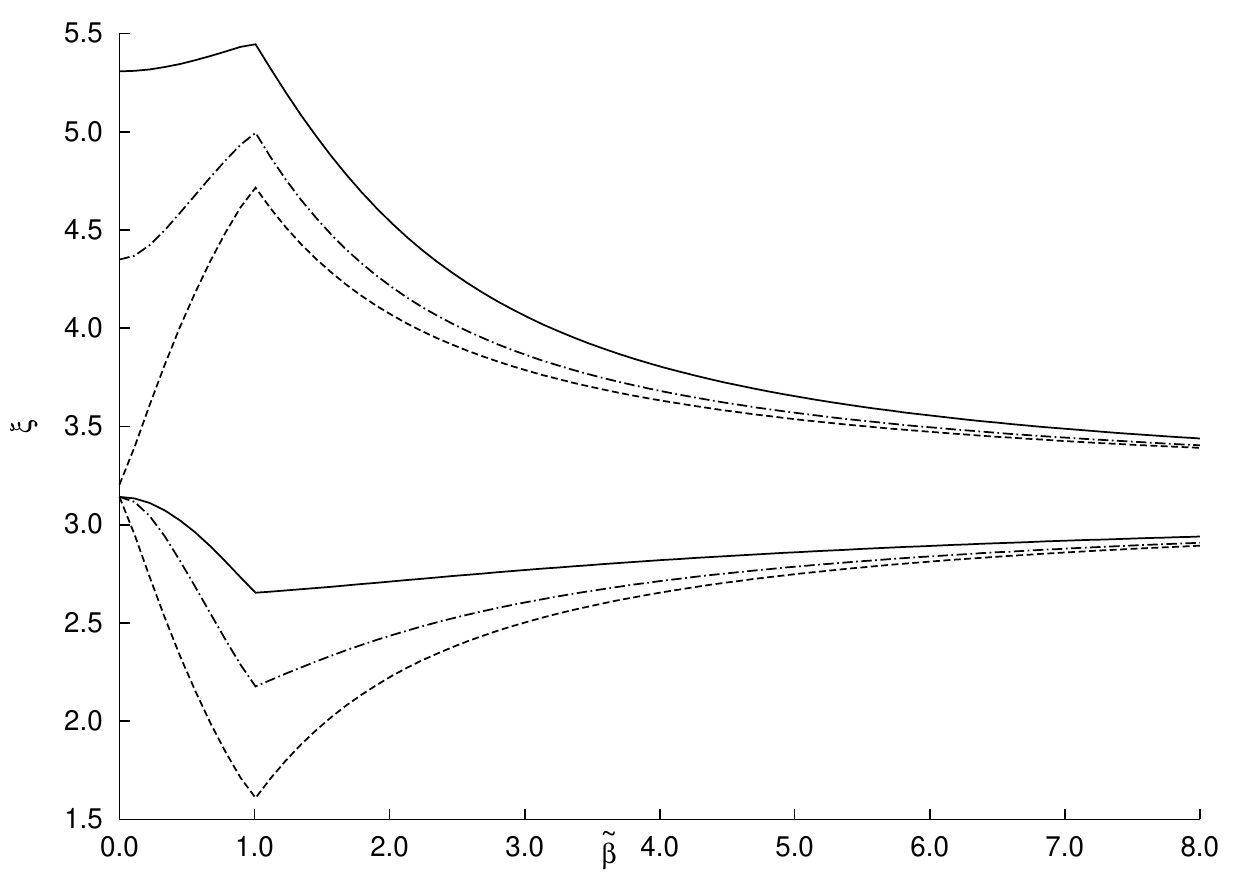}}
	\vspace{-0.3cm}
	\caption{\small{Dependence of a value $\xi=k_0L$ for the bottom and upper boundaries of the first (the upper Figure) and the second (the bottom Figure) forbidden zone on a parameter $\tilde{\beta}$ at different values of a parameter $p$: $p=10$ (solid line), $p=3$ (dash-dotted line) and $p=0.1$ (dashed line).}}
	\label{fig:U1U2U3}
\end{figure}

\section{Kronig-Penney model. Quantum wave impedance approach}
Two previous models, namely a Dirac comb and a $\delta-\delta'$ comb, containe singular potentials.
The Kronig-Penny model is devoid of this ``shortcoming'' and at the same time allows gettinng an exact solution. This model was introduced in 1931 \cite{Kronig_Penney:1931} and it consists of an infinite periodic set of rectangular potential barriers. The potentil energy in this model has a form
\begin{eqnarray}
U(x)=\sum_{n=-\infty}^\infty U_s(x+nL) , 
\end{eqnarray}
where 
\begin{eqnarray}
U_s(x)=\left\{\begin{array}{c}
0, \ \ 0<x\leq a\\
U_b, \ \ a<x\leq L
\end{array}\right.,
\end{eqnarray}
and $L=a+b$. The solution of the equation for a quantum wave impedance function \cite{Arx1:2020} in this case is as follows
\begin{eqnarray}
Z(x)=\left\{\begin{array}{c}
z_1\th\left[ik_1x+\phi_1\right], \ \ 0<x\leq a\\
z_2\th\left[\varkappa_2x+\phi_2\right], \ \ a<x\leq L
\end{array}\right.
\end{eqnarray}
with matching and periodic conditions
\begin{eqnarray}
Z(a-0)=Z(a+0),\quad Z(0)=Z(a+b),	
\end{eqnarray}
which means
\begin{eqnarray}\label{pm_cond_dd'}
z_1\th\left[ik_1a+\phi_1\right]&=&
z_2\th\left[\varkappa_2a+\phi_2\right],\nonumber\\
z_1\th\left[\phi_1\right]&=&
z_2\th\left[\varkappa_2(a+b)+\phi_2\right],
\end{eqnarray}
where $k_1=\sqrt{2mE}/\hbar$, $\varkappa_2=\sqrt{2m(U_b-E)}/\hbar$, $z_1=\sqrt{2E/m}$,
$z_2=\sqrt{2(E-U_b)/m}$.
Using (\ref{Z_periodic}) we get:
\begin{eqnarray}\label{disp_rel_KP_nf}
\frac{\ch\left[ik_1a+\phi_1\right]}{\ch\left[\phi_1\right]}
\frac{\ch\left[\varkappa_2(a+b)+\phi_2\right]}{\ch\left[\varkappa_2a+\phi_2\right]}=\exp(ikL).
\end{eqnarray}
It gives
\begin{eqnarray} 
&&\left[\frac{}{}\ch(ik_1a)+\sh(ik_1a)\th(\phi_1)\right]
\left[\frac{}{}\ch(\varkappa_2b)+\sh(\varkappa_2b)\right.\times\nonumber\\
&&\times\left.\frac{}{}\th(\varkappa_2a+\phi_2)\right]=
\ch(ik_1a)\ch(\varkappa_2b)+
\sh(ik_1a)\sh(\varkappa_2b)\times\nonumber\\
&&\times\left(\th(\phi_1)\th(\varkappa_2a+\phi_2)+\frac{\th(\phi_1)}
{\th(\varkappa_2b)}+
\frac{\th(\varkappa_2a+\phi_2)}{\th(ik_1a)}\right).
\end{eqnarray}
On the base of (\ref{pm_cond_dd'}) we get
\begin{eqnarray}
\th(\varkappa_2a+\phi_2)=\frac{z_1}{z_2}\frac{\th(ik_1a)+\th(\phi_1)}{1+\th(ik_1a)\th(\phi_1)},
\nonumber\\
\th(\phi_1)=\frac{z_2}{z_1}\frac{\th(\varkappa_2b)+\th(ik_1a+\phi_2)}{1+\th(\varkappa_2b)\th(ik_1a+\phi_2)}.\nonumber\\
\end{eqnarray}
Taking into account that $k_1$ and $\varkappa_2$ are real we obtain
\begin{eqnarray}
\Im\left(\th(\phi_1)\th(\varkappa_2a+\phi_2)+\frac{\th(\phi_1)}
{\th(\varkappa_2b)}+
\frac{\th(\varkappa_2a+\phi_2)}{\th(ik_1a)}\right)=\frac{z_1^2+z_2^2}{2z_1z_2}=\frac{k_1^2+\varkappa_2^2}{2k_1\varkappa_2}.
\end{eqnarray}
Taking a real part from both sides of an equation (\ref{disp_rel_KP_nf}) we get the well-known dispersion relation for a Kronig-Penney model:
\begin{eqnarray}\label{DR_KP}
\cos(k L)=\cos(k_1a)\ch(\varkappa_2b)+\frac{\varkappa_2^2-k_1^2}{2k_1\varkappa_2}\sin(k_1a)\sh(\varkappa_2b).
\end{eqnarray}

\section{Kronig-Penney model. Classical approach}
In the previous section we have shown how a quantum wave impedance approach can be applied for a calculation of a dispersion relation for a Kronig-Penney model. To understand the adventures of this approach we are going to compare it with the classical one based on the direct solving of a Sr\"{o}dinger equation. 

General solution of a Shr\"{o}dinger equation with a periodic potential accordingly to the Bloch-Floquet theorem can be shown as:
\begin{eqnarray}
\psi(x)=A_+u_k(x)\exp[ik x]+A_-u_{-k}(x)\exp[-ik x],
\end{eqnarray}
where
\begin{eqnarray}
u_I=A_+\exp[i(k_1-k)x]+A_-\exp[-i(k_1+k)x],\nonumber\\
u_{II}=B_+\exp[i(k_2-k)x]+B_-\exp[-i(k_2+k)x],
\end{eqnarray}
$k_1=\frac{\sqrt{2mE}}{\hbar}$ and $ik_2=\varkappa_2=\frac{\sqrt{2m(U_b-E)}}{\hbar}$ (as in the previous section). Periodic and matching conditions are
\begin{eqnarray}
u_I(0)=u_{II}(0),\quad u_I'(0)=u_{II}'(0),\nonumber\\
u_I(-b)=u_{II}(a),\quad u_I'(-b)=u_{II}'(a).
\end{eqnarray}
Thus, it gives the system of equations:
\begin{eqnarray}
&&A_++A_-=B_++B_-,\nonumber\\
&&A_+k_1-A_-k_1=B_+k_2-B_-k_2,\nonumber\\
&&A_+e^{-i(k_1-k)b}+A_-e^{i(k_1+k)b}=
B_+e^{i(k_2-k)a}+B_-e^{-i(k_2+k)a},\nonumber\\
&&A_+(k_1-k)e^{-i(k_1-k)b}-A_-(k_1+k)e^{i(k_1+k)b}=\nonumber\\&&=B_+(k_2-k)e^{i(k_2-k)a}-B_-(k_2+k)e^{-i(k_2+k)a}.
\end{eqnarray}
This system of equations can be represented in a matrix form
\begin{eqnarray}
\!\!\!\begin{pmatrix}
1 & 1 & -1 &-1 
\\
k_1 & -k_1 & -k_2 & k_2
\\ 
e^{-i\Delta_1b} & e^{i\Sigma_1b} & -e^{i\Delta_2a} & -e^{i\Sigma_2a}
\\
\Delta_1e^{-i\Delta_1b}& -\Sigma_1e^{i\Sigma_1b}&
-\Delta_2e^{i\Delta_2a}& \Sigma_2e^{-i\Sigma_2a}
\end{pmatrix}
\!\!
\begin{pmatrix}
A_+\\
A_-\\
B_+\\
B_-
\end{pmatrix}=
\begin{pmatrix}
0\\
0\\
0\\
0
\end{pmatrix}\!\!,
\end{eqnarray}
where $\Delta_1=k_1-k$, $\Delta_2=k_2-k$, $\Sigma_1=k_1+k$, $\Sigma_2=k_2+k$. 
From the condition $Det(M)=0$ and reminding that $ik_2=\varkappa_2$ we obtain the solution (\ref{DR_KP}).

In this approach the most routine and exhausting is the process of solving the last system of equations. This is due to the fact that calculating determinant of $4 \times 4$ size with further simplification of the obtained expression demands a lot of efforts.

\section{Kronig-Penney model. Transfer matrix app\-roach}
In this section we will solve the  Kronig-Penney model using a transfer matrix approach. Together with the two previous approaches it will give us the clear understanding of advantages and disadvantages of each approach.

Transfer matrix for an elementary cell ($0\leq x<L$) consists of 4 matrices, namely, $T=M_{I}I_{I\rightarrow II}M_{II}I_{II\rightarrow I}$, where
\begin{eqnarray}
I_{I\rightarrow II}=\frac{1}{2} 
\begin{pmatrix}
\frac{ik_1+\varkappa_2}{\varkappa_2}, & \frac{\varkappa_2-ik_1}{\varkappa_2} 
\\
\frac{\varkappa_2-ik_1}{\varkappa_2}, & \frac{ik_1+\varkappa_2}{\varkappa_2}
\end{pmatrix},
\!\!\!\quad
I_{II\rightarrow I}=\frac{1}{2} 
\begin{pmatrix}
\frac{\varkappa_2+ik_1}{ik_1} & \frac{ik_1-\varkappa_2}{ik_1} 
\\
\frac{ik_1-\varkappa_2}{ik_1} & \frac{\varkappa_2+ik_1}{ik_1}
\end{pmatrix}\!\!
\end{eqnarray}
describe a quantum wave function transferring through an interface of regions with a different potential energy. If a wave moves from a region I ($0 \leq x<a$) to a region II ($a \leq x<a+b$) we use $I_{I\rightarrow II}$ and otherwise $I_{II\rightarrow I}$.

Wave transferring inside regions $I$ and $II$ is described by matrices $M_I$ and $M_{II}$ 
\begin{eqnarray}
M_I=\begin{pmatrix}
e^{ik_1a} & 0 
\\
0 & e^{-ik_1a}
\end{pmatrix},\quad
M_{II}=\begin{pmatrix}
e^{\varkappa_2b} & 0 
\\
0 & e^{-\varkappa_2b}
\end{pmatrix}\!\!.
\end{eqnarray}
Thus, a transfer matrix for an elementary cell has a form
\begin{eqnarray}
\!\!\!\!\!T\!\!\!&=&\!\!\!\frac{1}{4} 
\begin{pmatrix}
e^{ik_1a} & 0 
\\
0 & e^{-ik_1a}
\end{pmatrix}
\begin{pmatrix}
\frac{ik_1+\varkappa_2}{\varkappa_2} & \frac{\varkappa_2-ik_1}{\varkappa_2} 
\\
\frac{\varkappa_2-ik_1}{\varkappa_2} & \frac{ik_1+\varkappa_2}{\varkappa_2}
\end{pmatrix}
\times\nonumber\\
\!\!\!&\times&\!\!\!
\begin{pmatrix}
e^{\varkappa_2b}& 0 
\\
0& e^{-\varkappa_2b}
\end{pmatrix}
\begin{pmatrix}
\frac{\varkappa_2+ik_1}{ik_1}& \frac{ik_1-\varkappa_2}{ik_1} 
\\
\frac{ik_1-\varkappa_2}{ik_1}& \frac{\varkappa_2+ik_1}{ik_1}
\end{pmatrix}=\nonumber\\
\!\!\!\!\!\!&=&\!\!\!\!\!\!\begin{pmatrix}
\!\!
\left(i\frac{k_1^2-\varkappa_2^2}{2k_1\varkappa_2}\sh[\varkappa_2b]\!+\!\ch[\varkappa_2b]\right)\!e^{ik_1a}&\!\!\!\!\!\!\!\!\!\!\!\!\!\! i\frac{k_1^2+\varkappa_2^2}{2k_1\varkappa_2}\sh[\varkappa_2b]e^{ik_1a}
\\
-i\frac{k_1^2+\varkappa_2^2}{2k_1\varkappa_2}\sh[\varkappa_2b]e^{-ik_1a}& \!\!\!\!\!\!\!\!\!\!\!\!\!\!\left(i\frac{\varkappa_2^2-k_1^2}{2k_1\varkappa_2}\sh[\varkappa_2b]+\ch[\varkappa_2b]\right)\!e^{-ik_1a}\!\!
\end{pmatrix}\!\!.\nonumber\\
\end{eqnarray}
To get the dispersion relation we use the following condition:
\begin{eqnarray}
\cos(kL)=\frac{1}{2}(T_{11}+T_{22}).
\end{eqnarray}
and as a result we obtain a formula (\ref{DR_KP}).

\section{Semi-infinite systems}
Up to this time we considered only perfectly periodic infinite systems.
But it is well-known that a violation of a lattice periodicity causes the appearance of energy levels in forbidden zones. In this section we consider one of the simplest cases of a periodicity violation, namely a presence of one side edge, which bears a class of semi-infinite systems. For the first time this problem was solved by Tamm \cite{Tamm:1932, Tamm:1933} long ago. And he has shown the existence of surface states in this system. Here we will show how to apply a quantum wave impedance approach for solving this model.

So let's consider a semi-infinite Dirac comb. The potential energy of this model can be depicted as:
\begin{eqnarray}
U(x)=U_E\theta(-x)+\alpha\sum_{n=1}^\infty \delta(x-na). 
\end{eqnarray}
The solution (for a region $0<x<a$) of an equation for a quantum wave impedance function \cite{Arx1:2020} with a  potential of periodically placed $\delta$-functions we have got in previous sections and it is  
\begin{eqnarray}
Z(x)=z_0\th[ik_0x+\phi]=
z_0\frac{\exp[ik_0x]-r_0\exp[-ik_0x]}{\exp[ik_0x]+r_0\exp[-ik_0x]},
\end{eqnarray}
where
\begin{eqnarray}\label{r0_k0k}
r_0=\exp[-2\phi]=\frac{\exp[i(k_0-k)L]-1}{1-\exp[-i(k_0+k)L]}.
\end{eqnarray}
In a scattering case, when $k$ is real, we find that
\begin{eqnarray}
r=\frac{Z(0)+z_E}{Z(0)-z_E}=
\frac{z_0(1-r_0)+z_E(1+r_0)}{z_0(1-r_0)-z_E(1+r_0)},
\end{eqnarray}
here $z_E=\sqrt{2(E-U_E)/m}$ is the characteristic impedance of a medium to the left of the point $x=0$.

It turns out that in such a system the bound states exist as well. To find them we have to solve the equation $Z(0)=-z_E$ for an imaginary $k$. This simple equation gives the next one
\begin{eqnarray}
z_0(1-r_0)=-z_E(1+r_0)
\end{eqnarray}
or, using the expression for $r_0$ (\ref{r0_k0k}), the following one
\begin{eqnarray}
\frac{z_0+z_E}{z_0-z_E}=\frac{\exp[i(k_0-k)L]-1}{1-\exp[-i(k_0+k)L]}.
\end{eqnarray}
After simple transformations we get:
\begin{eqnarray}
\exp[ik L]=\cos(k_0L)-iz_E/z_0\sin(k_0L) 
\end{eqnarray} 
or in the other notations
\begin{eqnarray}
\exp[ik L]=\cos(\xi)+\frac{\sqrt{s^2-\xi^2}}{\xi}\sin(\xi), 
\end{eqnarray}
where $\xi\!=\!k_0L$, $s\!=\!\sqrt{\varkappa_E^2L^2\!+\!\xi^2}=\sqrt{2mU_E}L/\hbar$, $\varkappa_E\!=\!\sqrt{2m(U_E\!-\!E)}/\hbar$.
A quasi wave-vector $k$ should be taken from a forbidden zone. It means that we can take $k$ in a form
\begin{eqnarray}\label{gm_lb_n}
k=i\lambda+\pi n/L, \qquad n=0,1,2,\ldots,
\end{eqnarray}
where $n$ numbers zones.
Finally it gives us the equation 
\begin{eqnarray}\label{SS_exp1}
(-1)^n\exp[-\lambda L]=\cos(\xi)+\frac{\sqrt{s^2-\xi^2}}{\xi}\sin(\xi)
\end{eqnarray}
which together with the dispersion relation (\ref{dr_DC}) in an appropriate form, namely
\begin{eqnarray}\label{SS_exp2}
(-1)^n\ch(\lambda L)=\cos(\xi)+\frac{p}{\xi}\sin(\xi)
\end{eqnarray}
allows getting an expression for finding energies of surface states:
\begin{eqnarray}\label{SS_exp}
\xi\ctg(\xi)=\frac{s^2}{2p}-\sqrt{s^2-\xi^2}.
\end{eqnarray}

\section{Semi-infinite system with a deformed edge}
In further papers one-dimensional surface state models were investigated precisely in a variety of cases (see, for example, \cite{Koutecky:1957, Koutecky:1960, Aerts:1960_1, Aerts:1960_2, Aerts:1960_3, Roy_Pandey:1988}). In this section we would like to consider the system which is a little complicated compare to the one from the previous section. So we deal with the 1D semi-infinite crystal in which the first cell is
deformed. This deformation is represented by a term $\eta\delta(x)$ in the potential energy of the studied system. Thus, we have a semi-infinite Dirac comb with a deformed edge and this model is closer to the real systems \cite{Lander_Gobeli_Morrison:1963,Marsh_Farnsworth:1964} than the previous one. This system was considered and solved in \cite{Steslicka_Wojciechowski:1966, Davison_Steslicka:1971}. But here we apply a quantum wave impedance approach for solving this model. So a potential energy has a following form
%So the potential energy of this system has a form:
\begin{eqnarray}
U(x)=U_0\theta(-x-a)+\eta\delta (x)+\sum_{n=1}^\infty \delta(x-na).
\end{eqnarray}
In a point $x=0$ we have to apply the matching condition \cite{Arx4:2020}:
\begin{eqnarray}
Z(0+)=Z(0-)-\frac{2i\eta}{\hbar}. 
\end{eqnarray}
It is easy to find $Z(0-)$ using the iterative formula for a quantum wave impedance calculation \cite{Arx3:2020}:
\begin{eqnarray}
Z(0-)=z_0\frac{-z_E\ch[ik_0L]+z_0\sh[ik_0L]}
{z_0\ch[ik_0L]-z_E\sh[ik_0L]}=-\frac{\hbar}{m}k_0\frac{\varkappa_E-k_0\tg(k_0L)}
{k_0+\varkappa_E\tg(k_0L)},
\end{eqnarray}
where $\varkappa_E=\frac{m}{i\hbar}z_E$, $k_0=\frac{m}{\hbar}z_0$.
In the previous section we have already found that $Z(0+)=z_0(1-r_0)/(1+r_0)$. 
Thus, we get a condition for finding surface states in a semi-infinite Dirac comb with a deformed oneside edge 
\begin{eqnarray}
-i\frac{\varkappa_E-k_0\tg(k_0L)}
{k_0+\varkappa_E\tg(k_0L)}-\frac{2im}{\hbar^2 k_0}\eta=
\frac{1-r_0}{1+r_0}.
\end{eqnarray}
Taking into account the expression for $r_0$ (\ref{r0_k0k}) and after simple trans\-formations one get:
\begin{eqnarray}
\exp[ik L]=\cos(k_0L)+\left(\frac{\varkappa_E-k_0\tg(k_0L)}
{k_0+\varkappa_E\tg(k_0L)}+\frac{2m\eta}{\hbar^2 k_0}\right)
\sin(k_0L).
\end{eqnarray}
or in the other notations
\begin{eqnarray}
\exp[ik L]=\cos(\xi)+\left(\frac{\varkappa_EL-\xi\tg(\xi)}
{\xi+\varkappa_EL\tg(\xi)}+\frac{p_\eta}{\xi}\right)
\sin(\xi).
\end{eqnarray}
Taking $k$ from the forbidden zone in the form $k=i\lambda+\pi n/a, n=0,1,2,\ldots$ as in the previous section we finally obtain the expression which relates $\lambda$ with $k_0, \varkappa_E, L, \eta$:
\begin{eqnarray}
\!\!\!(\!-1)^n\!\exp[-\lambda L]\!=\!\cos(\xi)+\left(\frac{\varkappa_EL-\xi\tg(\xi)}
{\xi+\varkappa_EL\tg(\xi)}+\frac{p_\eta}{\xi}\right)
\sin(\xi).
\end{eqnarray}
To simplify the further transformations we introduce the following notations:
\begin{eqnarray}
t=\frac{\varkappa_EL\!-\!\xi\tg(\xi)}
{\xi\!+\!\varkappa_EL\tg(\xi)},
\quad p_\eta=\frac{m\eta L}{\hbar^2},
\quad p_\alpha=\frac{m\alpha L}{\hbar^2}.
\end{eqnarray}
Combining both this expression and a dispersion relation for a Dirac comb in an appropriate form (\ref{SS_exp2})  we get
\begin{eqnarray}
&&\frac{1}{2}\frac{\left(\cos(\xi)\!+\!\left(t+2p_\eta\right)\sin(\xi)/\xi\right)^2+1}{\cos(\xi)\!+\!\left(t+2p_\beta\right)\sin(\xi)/\xi}=\cos(\xi)+p_\alpha\sin(\xi)/\xi.
\end{eqnarray}
After transformations this formula takes on a fairly simple form
\begin{eqnarray}
\xi\ctg(\xi)=\frac{\xi^2+t^2+2t(2p_\eta-p_\alpha) +4p_\eta(p_\eta-p_\alpha)}
{2p_\alpha}.
\end{eqnarray}
Taking into account that 
\begin{eqnarray}
t^2+\xi^2&=&\xi^2\left[\left(\frac{\varkappa_EL\!-\!\xi\tg(\xi)}
{\xi\!+\!\varkappa_EL\tg(\xi)}\right)^2+1\right]=\frac{\varkappa_E^2L^2+\xi^2}{\left(\sin(\xi)+\varkappa_EL/\xi\cos(\xi)\right)^2}
\end{eqnarray}
and
\begin{eqnarray}
t\!\!\!&=&\!\!\!\frac{\varkappa_EL\!-\!\xi\tan(\xi)}
{\xi\!+\!\varkappa_EL\tan(\xi)}\!=\!\frac{(\varkappa_E^2L^2-\xi^2)\tg(\xi)+\varkappa_EL\xi(1-\tg^2(\xi))}
{(\xi\!+\!\varkappa_EL\tg(\xi))^2}\!=\nonumber\\
&=&\frac{(\varkappa_E^2L^2-\xi^2)/(2\xi^2)\sin(2\xi)+\varkappa_EL/\xi(\cos(2\xi))}
{\left(\sin(\xi)+\varkappa_EL/\xi\cos(\xi)\right)^2}
\end{eqnarray}
we finally get the expression which determines energies of surface states in a studied model
\begin{eqnarray}
\xi\!\ctg(\xi)=2\frac{p_\eta}{p_\alpha}(p_\eta\!-\!p_\alpha)\!+\!\frac{\frac{\varkappa_E^2L^2\!+\!\xi^2}{2p_\alpha}\!+\!(2\eta\!-\!\alpha)\!\left[(\varkappa_E^2L^2\!-\!\xi^2)/(2\xi^2)\sin(2\xi)\!+\!\varkappa_EL/\xi(\cos(2\xi))\right]}{\left(\sin(\xi)+\varkappa_EL/\xi\cos(\xi)\right)^2}.\nonumber\\
\end{eqnarray}
This expression coincides with the one obtained earlier in the papers which were mentioned at the beginning of this section.
 
 \newpage

\section*{Conclusions}
 
 We managed  to reformulate the problem of study of infinite and semi-infinite periodic systems  in terms of a quantum wave impedance method. As a result we obtained an instrument of the consideration such systems which allow simplifying their investigation significantly. To demonstrate this we provided the comparision of three differnet approaches, namely classical approach, transfer matrix technique and a quantum impedance method for study of a Kronig-Penney model. Using quantum wave impedance approach it was also shown that in the semi-infinite systems the violation of a periodicity 
 leads to the formation of energy levels in band gaps. 
 
 The same conclusion concerning the violation of a periodicity was obtained in  \cite{Nelin:2009_2}. The author of that article also states that
 using the impedance concept makes it possible to
 obtain new results generalizing the wave properties of
 various crystal-like structures and that the band nature of crystal-like structures as well as the band gap conditions follow from the impedance model without
 Bloch’s theorem. He makes the conclusion that such an approach is distinguished
 by its universality, clarity, and possibility of using analogues from various fields of engineering.

The methods derived in this paper can be applied for a wide scope of different infinite and semi-infinite systems especially in a cese of their numerical study.

\renewcommand\baselinestretch{1.0}\selectfont
%\renewcommand{\bibname}{Bibliography} 

%\fancyhead[RE,LO]{\sl Bibliography}

\def\name{\vspace*{-0cm}\LARGE 
	%СПИСОК ВИКОРИСТАНИХ ДЖЕРЕЛ
	Bibliography\thispagestyle{empty}}
\addcontentsline{toc}{chapter}{Bibliography}

{\small

	\bibliographystyle{gost780u}
	%\bibliography{\figsfolder full,add}
	\bibliography{full.bib}

\begin{thebibliography}{10}
\def\selectlanguageifdefined#1{
\expandafter\ifx\csname date#1\endcsname\relax
\else\language\csname l@#1\endcsname\fi}
\ifx\undefined\url\def\url#1{{\small #1}}\else\fi
\ifx\undefined\BibUrl\def\BibUrl#1{\url{#1}}\else\fi
\ifx\undefined\BibAnnote\long\def\BibAnnote#1{}\else\fi
\ifx\undefined\BibEmph\def\BibEmph#1{\emph{#1}}\else\fi

\bibitem{Hague_Hague_Khan:1990}
\selectlanguageifdefined{english}
\BibEmph{Haque~A.} Energyband calculation for periodic potential structure
  using quantum mechanical impedance~/ A.~Haque, M.~Haque, M.~R.~Khan~/$\!$/ J.
  Appl. Phys. "---
\newblock 1990. "---
\newblock Vol.~68, No.~4. "---
\newblock P.~1661--1664.

\bibitem{Nelin:2004}
\selectlanguageifdefined{english}
\BibEmph{Nelin~E.~A.} Simulation and improvement of the selectivity of
  crystal-like structures~/ E.~A.~Nelin~/$\!$/ Tech. Phys. "---
\newblock 2004. "---
\newblock Vol.~49, No.~11. "---
\newblock P.~1464--1468.

\bibitem{Nelin:2005}
\selectlanguageifdefined{english}
\BibEmph{Nelin~E.~A.} Edge apodization of crystal-like structures~/
  E.~A.~Nelin~/$\!$/ Tech. Phys. "---
\newblock 2005. "---
\newblock Vol.~50, No.~11. "---
\newblock P.~1511--1512.

\bibitem{Nelin:2006}
\selectlanguageifdefined{english}
\BibEmph{Nelin~E.~A.} Phase apodization of crystal-like structures~/
  E.~A.~Nelin~/$\!$/ Tech. Phys. "---
\newblock 2006. "---
\newblock Vol.~51, No.~8. "---
\newblock P.~1101--1103.

\bibitem{Nelin:2009_2}
\selectlanguageifdefined{english}
\BibEmph{Nelin~E.~A.} {I}mpedance {C}haracteristics of {C}rystal-like
  {S}tructures~/ E.~A.~Nelin~/$\!$/ Tech. Phys. "---
\newblock 2009. "---
\newblock Vol.~54, No.~7. "---
\newblock P.~953--957.

\bibitem{Khatyan_Gindikina_Nelin:2015}
\selectlanguageifdefined{english}
\BibEmph{Khatyan~D.~V.} Semiconductor superlattice zone diagram formation~/
  D.~V.~Khatyan, M.~A.~Gindikina, E.~A.~Nelin~/$\!$/ Visn. {NTUU} {KPI} {S}er.
  - {R}adiotekh. {R}adioaparatobud. "---
\newblock 2015. "---
\newblock Vol.~62. "---
\newblock P.~100--107.

\bibitem{Nazarko_etall:2009}
\selectlanguageifdefined{english}
\BibEmph{Nazarko~A.I.} Increasing of zone selectivity of electromagnetic
  crystals~/ A.I.~Nazarko, U.~F.~Tymofeeva, V.~I.~Nelin, N. A.~Popsuy~/$\!$/
  Tekhn. Konstr. Elektr. Appar. "---
\newblock 2009. "---
\newblock Vol.~6. "---
\newblock P.~38--41.

\bibitem{Gindikina_Zinger_Nelin:2015}
\selectlanguageifdefined{english}
\BibEmph{Gindikina~M.~A.} Zone diagram formation of photon and phonon
  crystals~/ M.~A.~Gindikina, Y.~L.~Zinger, E.~A.~Nelin~/$\!$/ Visn. {NTUU}
  {KPI} {S}er. - {R}adiotekh. {R}adioaparatobud. "---
\newblock 2015. "---
\newblock Vol.~63. "---
\newblock P.~119--126.

\bibitem{Nelin:2007_1}
\selectlanguageifdefined{english}
\BibEmph{Nelin~E.~A.} Narrowband filters based on one-barrier crystallike
  structures~/ E.~A.~Nelin~/$\!$/ Tech. Phys. "---
\newblock 2007. "---
\newblock Vol.~52, No.~9. "---
\newblock P.~1222--1224.

\bibitem{Nelin_Nazarko:2012}
\selectlanguageifdefined{english}
\BibEmph{Nelin~E.A}. Resonance and band filtration on the basis of twophase
  crystallike structures~/ E.A~Nelin, A.~I.~Nazarko~/$\!$/ Tech. Phys. "---
\newblock 2012. "---
\newblock Vol.~57, No.~10. "---
\newblock P.~1449--1452.

\bibitem{Babushkin_Nelin:2011}
\selectlanguageifdefined{english}
\BibEmph{Vodolazka~M.~V.} Resonance filtration by two-phase resonators~/
  M.~V.~Vodolazka, A.~P.~Tolstenkova, Nelin~E.~A.~/$\!$/ Visn. {NTUU} {KPI}
  {S}er. - {R}adiotekh. {R}adioaparatobud. "---
\newblock 2014. "---
\newblock Vol.~57. "---
\newblock P.~113--120.

\bibitem{Nazarko_etall:2011}
\selectlanguageifdefined{english}
Electromagnetic crystals based on lowimpedance inhomogeneities~/ A.~I.~Nazarko,
  E.~A.~Nelin, V.~I.~Popsui, Yu.~F.~Timofeeva~/$\!$/ Tech. Phys. "---
\newblock 2011. "---
\newblock Vol.~56, No.~5. "---
\newblock P.~728--730.

\bibitem{Nelin:2012}
\selectlanguageifdefined{english}
\BibEmph{Nelin~E.~A.} Two phase crystal like structures~/ E.~A.~Nelin~/$\!$/
  Tech. Phys. "---
\newblock 2012. "---
\newblock Vol.~57, No.~1. "---
\newblock P.~59--62.

\bibitem{Nazarko_etall:2011_1}
\selectlanguageifdefined{english}
Two-phase electromagnetic crystal~/ A.~I.~Nazarko, E.~A.~Nelin, V.~I.~Popsui,
  Yu.~F.~Timofeeva~/$\!$/ Tech. Phys. Lett. "---
\newblock 2011. "---
\newblock Vol.~37, No.~2. "---
\newblock P.~185--187.

\bibitem{Nazarko_Timofeeva_Nelin:2010}
\selectlanguageifdefined{english}
\BibEmph{Nazarko~A.~I.} Interaction of electromagnetic field with the
  inhomogeneities of electromagnetic crystals~/ A.~I.~Nazarko,
  Yu.~F.~Timofeeva, E.~A.~Nelin~/$\!$/ Visn. {NTUU} {KPI} {S}er. - {R}adiotekh.
  {R}adioaparatobud. "---
\newblock 2010. "---
\newblock Vol.~41. "---
\newblock P.~65--68.

\bibitem{Nelin_Zinher_Popsui:2017}
\selectlanguageifdefined{english}
\BibEmph{Nelin~E.~A.} Combined electromagnetocrystalline inhomogeneities~/
  E.~A.~Nelin, Ya.~L.~Zinher, Popsui~V.~I.~/$\!$/ Visn. {NTUU} {KPI} {S}er. -
  {R}adiotekh. {R}adioaparatobud. "---
\newblock 2017. "---
\newblock Vol.~71. "---
\newblock P.~46--51.

\bibitem{Nelin:2004_1}
\selectlanguageifdefined{english}
\BibEmph{Nelin~E.A.} Simulation and apodization of crystallike structures~/
  E.A.~Nelin~/$\!$/ Radioelecton. Commun. Syst. "---
\newblock 2004. "---
\newblock Vol.~47, No.~7. "---
\newblock P.~16--22.

\bibitem{Nazarko_etall:2015}
\selectlanguageifdefined{english}
Input impedance characteristics of microstrip structures~/ A.~I.~Nazarko,
  M.~V.~Vodolazka, P.~S.~Bidenko, E.~A.~Nelin~/$\!$/ Visn. {NTUU} {KPI} {S}er.
  - {R}adiotekh. {R}adioaparatobud. "---
\newblock 2015. "---
\newblock Vol.~61. "---
\newblock P.~72--81.

\bibitem{Ashcroft_Mermin:1976}
\selectlanguageifdefined{english}
\BibEmph{Ashcroft~N.~W.}~/ N.~W.~Ashcroft, N.~D.~Mermin~/$\!$/ Solid State
  Physics. "---
\newblock Saunders College Publishing, 1976. "---
\newblock P.~848.

\bibitem{Floquet:1883}
\selectlanguageifdefined{english}
\BibEmph{Floquet~G.} Sur les {\'{e}}quations diff{\'{e}}rentielles
  lin{\'{e}}aires {\`{a}} coefficients p{\'{e}}riodiques~/ G.~Floquet~/$\!$/
  Annales scientifiques de l{'{\'{E}}}cole Normale Sup{\'{e}}rieure. "---
\newblock 1883. "---
\newblock Vol.~12. "---
\newblock P.~47--88.

\bibitem{Bloch:1929}
\selectlanguageifdefined{english}
\BibEmph{Bloch~F.} {\"{U}}ber die quantenmechanik der elektronen in
  kristallgittern~/ F.~Bloch~/$\!$/ Zeitschrift f{\"{u}}r Physik (in German).
  Springer Science and Business Media LLC. "---
\newblock 1929. "---
\newblock Vol.~52, No.~7.

\bibitem{Arx1:2020}
\selectlanguageifdefined{english}
\BibEmph{Hryhorchak~O.~I.} Reformulation of a transmission and reflection
  problems in terms of a quantum wave impedance function~/
  O.~I.~Hryhorchak~/$\!$/ arXiv:2010.04682. "---
\newblock 2020. "---
\newblock P.~1--14.

\bibitem{Arx2:2020}
\selectlanguageifdefined{english}
\BibEmph{Hryhorchak~O.~I.} An application of a quantum wave impedance approach
  for solving a nonsymmetric single well problem~/ O.~I.~Hryhorchak~/$\!$/
  arXiv:2010.05583. "---
\newblock 2020. "---
\newblock P.~1--10.

\bibitem{Arx4:2020}
\selectlanguageifdefined{english}
\BibEmph{Hryhorchak~O.~I.} Application of a quantum wave impedance method for
  zero-range singular potentials~/ O.~I.~Hryhorchak~/$\!$/ arXiv:2010.06930.
  "---
\newblock 2020. "---
\newblock P.~1--17.

\bibitem{Cordoba:1989}
\selectlanguageifdefined{english}
\BibEmph{C\'{o}rdoba~A.} {D}irac combs~/ A.~C\'{o}rdoba~/$\!$/ Lett. in Math.
  Phys. "---
\newblock 2011. "---
\newblock Vol.~17. "---
\newblock P.~191--196.

\bibitem{Gadella:2019}
\selectlanguageifdefined{english}
Band spectra of periodic hybrid $\delta-\delta'$ structures~/ M.~Gadella,
  J.~M.~Mateos~Guilarte, J.~M.~Mu{\=n}oz-Casta{\=n}eda [et~al.]~/$\!$/
  arXiv:1909.08603. "---
\newblock 2019. "---
\newblock P.~1--28.

\bibitem{Kronig_Penney:1931}
\selectlanguageifdefined{english}
\BibEmph{Kronig~R. de~L.} Quantum mechanics of electrons in crystal lattices~/
  R.~de~L.~Kronig, W.~G.~Penney~/$\!$/ Proc. R. Soc. Lond. "---
\newblock 1931. "---
\newblock Vol.~130. "---
\newblock P.~499-513.

\bibitem{Tamm:1932}
\selectlanguageifdefined{english}
\BibEmph{Tamm~I.} {\"{U}}ber eine m{\"{o}}gliche art der elektronenbindung an
  kristalloberfl{\"{a}}chen~/ I.~Tamm~/$\!$/ Zeitschrift f{\"{u}}r Physik. "---
\newblock 1933. "---
\newblock Vol.~76, No.~11. "---
\newblock P.~849--850.

\bibitem{Tamm:1933}
\selectlanguageifdefined{english}
\BibEmph{Tamm~I.} A possible binding of the electrons on a crystal surface~/
  I.~Tamm~/$\!$/ Zh. Eksp. Teor. Fiz. "---
\newblock 1933. "---
\newblock Vol.~3. "---
\newblock P.~34--43.

\bibitem{Koutecky:1957}
\selectlanguageifdefined{english}
\BibEmph{Kouteck{\'{y}}~Ja.} Contribution to the theory of the surface
  electronic states in the one-electron approximation~/
  Ja.~Kouteck{\'{y}}~/$\!$/ Phys. Rev. "---
\newblock 1957. "---
\newblock Vol.~108, No.~1. "---
\newblock P.~13--18.

\bibitem{Koutecky:1960}
\selectlanguageifdefined{english}
\BibEmph{Kouteck{\'{y}}~J.} On the theory of surface states~/
  J.~Kouteck{\'{y}}~/$\!$/ J. Phys. Chem. Solids. "---
\newblock 1960. "---
\newblock Vol.~14. "---
\newblock P.~233--240.

\bibitem{Aerts:1960_1}
\selectlanguageifdefined{english}
\BibEmph{Aerts~E.} Sufrace states of one-dimensional crystals ({I})~/
  E.~Aerts~/$\!$/ Physica. "---
\newblock 1960. "---
\newblock Vol.~26. "---
\newblock P.~1047--1056.

\bibitem{Aerts:1960_2}
\selectlanguageifdefined{english}
\BibEmph{Aerts~E.} Sufrace states of one-dimensional crystals ({II})~/
  E.~Aerts~/$\!$/ Physica. "---
\newblock 1960. "---
\newblock Vol.~26. "---
\newblock P.~1057--1062.

\bibitem{Aerts:1960_3}
\selectlanguageifdefined{english}
\BibEmph{Aerts~E.} Sufrace states of one-dimensional crystals ({III})~/
  E.~Aerts~/$\!$/ Physica. "---
\newblock 1960. "---
\newblock Vol.~26. "---
\newblock P.~1063--1072.

\bibitem{Roy_Pandey:1988}
\selectlanguageifdefined{english}
\BibEmph{Roy~C.~L.} Surface states of a one-dimensional finite crystal with
  deformations at the surfaces~/ C.~L.~Roy, J.~S.~Pandey~/$\!$/ J. Phys. Chem.
  Solids. "---
\newblock 1988. "---
\newblock Vol.~49, No.~11. "---
\newblock P.~1373--1376.

\bibitem{Lander_Gobeli_Morrison:1963}
\selectlanguageifdefined{english}
\BibEmph{Lander~J.~J.} Structural properties of cleaved silicon and germanium
  surfaces~/ J.~J.~Lander, G.~W.~Gobeli, J.~Morrison~/$\!$/ J. Appl. Phys. "---
\newblock 1963. "---
\newblock Vol.~34, No.~8. "---
\newblock P.~2298--2306.

\bibitem{Marsh_Farnsworth:1964}
\selectlanguageifdefined{english}
\BibEmph{Marsh~J.~B.} Low-energy electron diffraction studies of (100) and
  (111) surfaces of semiconducting diamond~/ J.~B.~Marsh,
  H.~E.~Farnsworth~/$\!$/ Surf. Sci. "---
\newblock 1964. "---
\newblock Vol.~1, No.~1. "---
\newblock P.~3--21.

\bibitem{Steslicka_Wojciechowski:1966}
\selectlanguageifdefined{english}
\BibEmph{St\c{e}{\'s}licka~M.} Surface states of a deformed one-dimensional
  crystal~/ M.~St\c{e}{\'s}licka, K.~F.~Wojciechowski~/$\!$/ Physica. "---
\newblock 1966. "---
\newblock Vol.~32. "---
\newblock P.~1274--1282.

\bibitem{Davison_Steslicka:1971}
\selectlanguageifdefined{english}
\BibEmph{Davison~S.~G.} Relativistic treatment of localized states. a review~/
  S.~G.~Davison, M.~St\c{e}{\'s}licka~/$\!$/ Int. Journ. of Quant. Chem. "---
\newblock 1971. "---
\newblock Vol.~4. "---
\newblock P.~445--453.

\bibitem{Arx3:2020}
\selectlanguageifdefined{english}
\BibEmph{Hryhorchak~O.~I.} Quantum wave impedance calculation for an arbitrary
  piesewise constant potential~/ O.~I.~Hryhorchak~/$\!$/ arXiv:2010.06263. "---
\newblock 2020. "---
\newblock P.~1--11.

\end{thebibliography}
	% insert this in the bbl after	 \begin{thebibliography}{}: \interlinepenalty=10000
	
}

\newpage

\end{document}